\pdfoutput=1
\documentclass[twoside,twocolumn]{article}

\usepackage{blindtext} 

\usepackage[sc]{mathpazo} 
\usepackage[T1]{fontenc} 
\usepackage{microtype} 

\usepackage[english]{babel} 

\usepackage[hmarginratio=1:1,top=32mm,columnsep=20pt]{geometry} 
\usepackage[hang, small,labelfont=bf,up,textfont=it,up]{caption} 
\usepackage{booktabs} 

\usepackage{lettrine} 

\usepackage{enumitem} 
\setlist[itemize]{noitemsep} 

\usepackage{abstract} 

\usepackage{titlesec} 
\renewcommand\thesection{\Roman{section}} 
\renewcommand\thesubsection{\roman{subsection}} 
\titleformat{\section}[block]{\large\scshape\centering}{\thesection.}{1em}{} 
\titleformat{\subsection}[block]{\large}{\thesubsection.}{1em}{} 

\usepackage{fancyhdr} 

\usepackage{titling} 

\usepackage{hyperref} 
\usepackage{authblk}
\usepackage{graphicx}
\usepackage{listings}
\usepackage{amsfonts}
\usepackage{makecell}
\usepackage{array}

\pretitle{\begin{center}\Huge\bfseries} 
\posttitle{\end{center}} 
\title{A heuristic extending the Squarified  treemapping algorithm} 
	
\author[1]{Antonio Cesarano\thanks{a.cesarano19@studenti.unisa.it}}
\author[1]{Filomena Ferrucci\thanks{fferrucci@unisa.it}}
\author[2]{Mario Torre\thanks{neugens@redhat.com}}

\affil[1]{Department of Computer Science, University of Salerno}
\affil[2]{Red Hat GmbH}

\date{\today} 
\renewcommand{\maketitlehookd}{%
\begin{abstract}
A heuristic extending the Squarified Treemap technique for the representation of hierarchical information as treemaps is presented. The original technique gives high quality treemap views, since items are laid out with rectangles that approximate squares, allowing easy comparison and selection operations. New key steps, with a low computational impact, have been introduced to yield treemaps with even better aspect ratios and higher homogeneity among items.
\end{abstract}
}

\begin{document}

\maketitle


In this paper, we present a heuristic extending the Squarified Treemap technique, which aims to lay out hierarchical information by treemap having square-like rectangles. Before explaining how such technique has been improved, an overview about the Squarified procedure is given. Comparison tests and discussions about the obtained results will close the paper.

\section{Introduction}

The Big Data era \cite{bigdata} is driving IT world towards visualization and computation of large amount of data, often structured in a hierarchical way, or more simply, as trees \cite{treemap26}. There are plenty of examples for such kind of structure, like large-scale simulations, high volume statistics and business data, or work breakdown structures of complex software projects \cite{treemap25}, and as many ways to present them graphically. One of the most used techniques for these data type structures is treemapping, introduced by Ben Shneiderman \cite{treemap2, treemap3}. A treemap is a technique for displaying hierarchical information through nested rectangles having size proportionate to data's weight. They appear compact and intuitive, giving to the viewer all the needed information immediately. There are lots of treemapping algorithm in literature and one of the most common techniques adopted is the Squarified Treemap, which aims to produce treemaps with square-like rectangles, making selection and comparison operations much easier and effective.\\
\\
In this paper, we present a heuristic that allows to extend the Squarified algorithm, which introduces the direction concept to the squarification process, yielding rectangles with an even better aspect ratio. We will call the extended version Squarified+. Moreover, we present some comparison tests to assess empirically the improvement of aspect ratio and homogeneity factors. This work has been developed during the Erasmus Traineeship period at Red Hat, and the Squarified+ algorithm has been applied to the heap analyzer module of Thermostat \cite{thermostat10} to show the heap memory composition allocated by JVMs.


\section{Background}
The problem of displaying hierarchical data can be faced with many different approaches, depending on the requirements that have to be satisfied. One of the first methods developed for this purpose was the node-edge diagram \cite{tree5, tree7, treemap17}, able to represent in the most faithful way a hierarchical structure. Since then, several techniques have been presented to improve its effectiveness \cite{tree8, tree6, treemap23}, as well as make graphic user interfaces more attractive \cite{hierarchiy4, treemap18}. However, this has not solved all the issues due to the intrinsic nature of trees \cite{treemap21}. We could think about the inability of using all the available space of canvases, in fact there is always a lot of wasted unused space around the root node, or the necessity of scrolling down the view to find leaf nodes in a deep structure \cite{tree27}, losing sight of higher nodes. Furthermore, it is very difficult individuating node-ancestors relationships among all the tree's edges in a dense structure.\\
\\
The treemapping technique resolves these problems, providing a compact view that makes full use of free available space and is able to give immediate details about nodes as well as general information about the overall structure. This is due to the nesting \cite{treemap9, beamtree11, treemap12} technique recursively applied on the tree that allows to nest many smaller treemaps into the parent node, to emphasize parent-child and sibling relationships, answering to questions like "Which is the biggest node? And the smallest one? Which node has more children?" \cite{treemap9}.\\
Treemaps work on three main factors when they lay out a tree: the order used to place rectangles into the view, the aspect ratio of generated rectangles, and the view stability, which is the items' propensity to maintain the position when the size varies due to a change of items' weight.\\
\\
In literature, we can find many treemapping techniques, each of them trying to optimize or emphasize one or more key factors; for example, the Hilbert and Moore\cite{treemap15} aims to improve treemaps' stability, ordered and quantum treemaps \cite{treemap13, treemap16} address the ordering factor, while Squarified optimizes the aspect ratio and spiral treemaps emphasize changes to the hierarchical structure. Moreover, alternative treemap methods have been developed, using circular \cite{treemap14} or polygonal shapes \cite{treemap18, treemap20}, achieving to fit non-rectangular canvases and providing a more pleasing way to present data \cite{treemap19}. Even if such kind of diagram could appear more attractive for newest graphic interfaces, the Squarified treemap is still one of the commonest techniques adopted in computer science field and not \cite{treemap26}.

\section{Squarified treemap - State of the art}
\begin{figure}[h]
	
	\centering
	\includegraphics[width=1\linewidth]{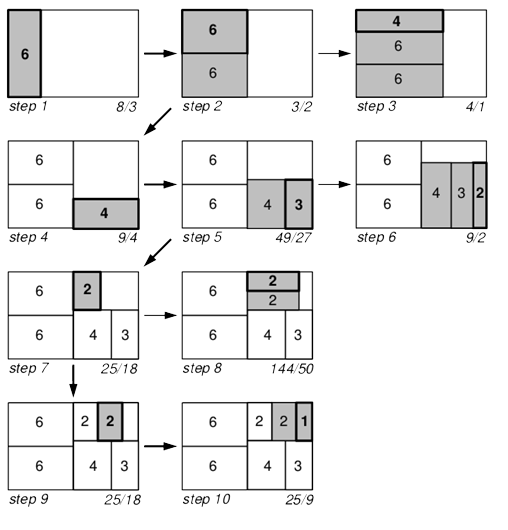}
	\caption{Example of Squarified algorithm \label{fig:squarified}}
\end{figure}

The Squarified method relies principally on the concept of row \cite{treemap24}, a set of items arranged adjacently in the parent container for which the aspect ratio is no longer improvable. Its goal is to calculate the best rows as possible, to avoid the generation of thin and elongated rectangle, making comparison and selection operations more effectiveness. The algorithm uses a global Rectangle datatype that we will call \textit{R}, representing the available space where items will be drawn, and takes in input a list of real numbers that are the item's weights, the current row, and a value w used to process the current row. Main steps are performed by the squarify function that recursively processes all given items. When an item is processed, a decision is taken between two alternatives. Either the rectangle is added to the current row, or the current row is fixed and a new row is initialized in the remaining space of the available area. This decision depends only on whether adding a rectangle to the row will improve the layout of the current row or not. The worst function is responsible for the row goodness, because it returns the aspect ratio values for the most elongated and skinny rectangle in the current row. In this way, it is possible to compare the current row quality with and without a new item. Moreover, the function width returns the smaller side of the available area of \textit{R}; squarify uses it because arranging items on the smaller side gives more chance to yield squared rectangles. Since this is a greedy algorithm, the input list must be ordered before starting.\\
\newline
Assuming to have a rectangle \textit{R} with width 6 and height 4, and a set of item $E = [6, 6, 4, 3, 2, 2,1]$. The treemapping process performed by the Squarified technique is summarized in the Figure ~\ref{fig:squarified}. 

\renewcommand{\figurename}{Figure}

We provide an extension for such process, introducing two further steps per row, having a minimal computational impact, that allow to:
\begin{itemize}
	\item{Improve the aspect ratio of treemap's items further.}
	\item{Improve the homogeneity factor among rectangles.}
\end{itemize}

\section{Squarified+}
Building an optimum treemap for a set of item falls into the NP class. This is why many treemapping techniques use a greedy approach, which approximates an optimum solution. However, there could be a way to improve their results to obtain an even better treemap that optimize a primary criterion, like ordering, aspect ratio or stability. The Squarified+ algorithm aims to extend the Squarified method by introducing a new key concept called direction that affects the way rows are arranged into the canvas. There are two main direction: left-to-right and top-to-bottom. Obviously, inverting each of them the result will not change.\\
When a new row is created, we calculate the current direction by checking the smaller side of the available area of \textit{R}, similarly to what is done in the original algorithm. The direction assumes the value top-to-bottom if the smaller side is the height, vice versa it turns into left-to-right. In this way, it is more probable to create squared rectangles. Once the squarefy function has calculated the best row to lay out, instead of fixing it in \textit{R}, we clone the current row to recalculate its items, drawing them in the opposite direction. This step gives the chance to improve items' aspect ratio, because more space is given to the row and elongated rectangles can be arranged in a better manner. Then, the current row and the alternative one are compared and the best one is chosen to be fixed.\\
\\
We present below this approach with a graphic example (Figure \ref{fig:squarified+}) that recalls the above figure for Squarified treemap, followed by a detailed description of all steps and changes.

\subsection{Example}
\begin{figure}[h]
	\centering
	\includegraphics[width=1\linewidth]{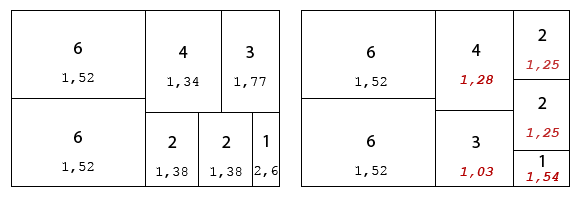}
	\caption{Differences between Squarified and Squarified+ executions	\label{fig:results}}
\end{figure}

\begin{figure}[h]
	\centering
	\includegraphics[width=1\linewidth]{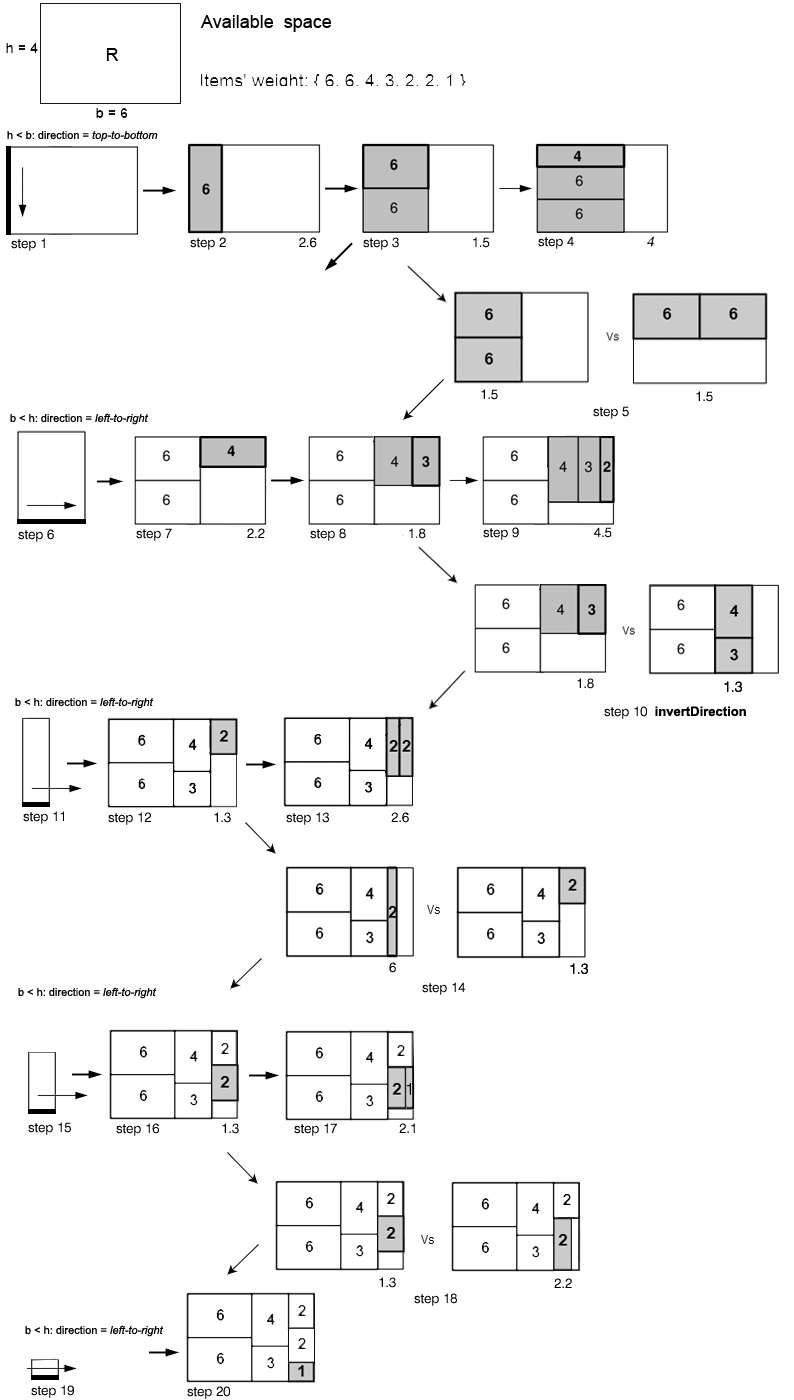}
	\caption{Squarified+ execution example \label{fig:squarified+}}
\end{figure}

Assuming to have the same input of the previous example, with a rectangle \textit{R} of 6x4, and a set of items E, already ordered. The first step consists of deciding the drawing direction that will be used to lay out the current row. The width is greater than the height, hence items will be arranged in a  top-to-bottom manner. Successive steps (2, 3, 4) regard the current row building process, adding items to the current row until no more elements can improve the global aspect ratio. This process is exactly the same as Squarified technique. Then, in the step 5, an alternative row is instantiated using the same items, but it is drawn in the opposite direction, which is left-to-right. Both configurations give an aspect ratio of 1.5 so we decide to don't change the direction, fixing the current row.\\
The new free area of \textit{R} is recalculated, and the direction is updated in the step 6, giving a left-to-right value. Again, in steps 7, 8, 9 the new current row is calculated and, in the step 10, an alternative row has been built in the opposite direction. This time, the alternative row gives a better aspect ratio than the original one, due to a wider space for drawing the rectangle with area 3, thus the alternative row is fixed. From this step, all rectangles will have a better aspect ratio in comparison to the original algorithm, even if in this example no more inversion of direction will be performed. Remaining rectangles are processed repeating these steps until no more items have to be drawn in the final treemap. Figure \ref{fig:results} compares a Squarified treemap with the corresponding Squarified+ one, highlighting improved aspect ratios.

\renewcommand{\figurename}{Figure}

\subsection{Algorithm}
Following the above example, we present now the extended algorithm Squarified+ for hierarchical information representation. As well as its original version, the squarefy function takes in input a list of reals, representing the nodes' weights, the current row, and a value w, which is the size of the side along which the current row will be arranged. After that the best row has been individuated, the function validate is responsible for building the alternative row, cloning the current one and processing it in the opposite direction, to evaluate the best choice. It relies on the \textit{smallerSide} and \textit{biggerSide} functions, that respectively give the smallest and the biggest side of the free space of the canvas \textit{R}, hold as a global variable. Furthermore, the function invertDirection, to switch the direction value, and the function improve are provided. The latter one allows evaluating which row, between the current one and the alternative one, gives the best output in terms of aspect ratio, that is rectangle more similar to squares. Thus, the alternative row improves the original one if the following Boolean expression becomes true: $ \frac {|actualAR - 1|} {|alternativeAR - 1|} > 1 $.
	
\renewcommand{\figurename}{Figure}

A true value result means that the worst rectangle in the alternative row is less thin and elongated than the worst rectangle in the current row. Thus, the alternative row improves the aspect ratio quality and the current direction value switches, vice versa, the current direction is keep. Using the direction value, the chosen row is fixed by the layoutrow function. The following piece of pseudocode illustrates where the extension has been applied to the Squarified technique:

\lstset{
	basicstyle=\ttfamily,
	keywordstyle=\bfseries,
	showstringspaces=false,
	morekeywords={validate, smallerSide, biggerSide, improve, invertDirection}
}
\vspace{0.5cm}
\renewcommand{\lstlistingname}{Code}
\begin{lstlisting}[frame=none, basicstyle=\footnotesize]
procedure squarify( list <real> children,
   list<real > row ,  real w)
  begin
    realC = head(children);
    if(worst(row, w)<= worst(row++[c],w)) 
      squarify(tail(children), row++[c], w)
    else
      validate(row);
      squarify(children,[],smallerSide());
    fi
  end
\end{lstlisting}
\vspace{0.5cm}

While the following pseudocode illustrates how the drawing direction is switched due to the evaluation between the current row and the alternative row, built on the bigger side of the available area of \textit{R}.

\vspace{0.5cm}
\renewcommand{\lstlistingname}{Code}
\begin{lstlisting}[frame=none, basicstyle=\footnotesize]
procedure validate( list<real> row )
  begin
   real actualAR=worst(row,smallerSide())   
   real alternativeAR=worst(row,biggerSide())
   if (improve(actualAR, alternativeAR)) then
     invertDirection()
   layoutrow(row)
  end
\end{lstlisting}

\section{Computational Cost Assessment}
One of the main goal of this work is to extend the Squarified algorithm without adding any significant computational cost. For this reason, we are going to analyze the effort needed to extend the Squarified technique by our heuristic, evaluating the computational cost of all introduced functions. 
\\
BiggerSide and smallerSide functions consist of a single ternary instruction that run in constant time, which is $\theta(1)$. The invertDirection function switches a variable between two states, costing  $\theta(1)$. The improve function evaluates which one from the given values in input is closer to 1, giving a computational cost of $\theta(1)$.
\\
The validate function is a little bit more complex, since it englobes all above functions and recall two times the worst method, which has an higher complexity in comparison to functions we have analyzed until now. Assuming we are processing a list of n nodes, the validate function is called every time a row has to be fixed in the canvas, that can be n times. Every invocation corresponds to two worst invocations. The size of the input of this method depends on how many nodes have still being processed. When the treemap starts processing items, there could be n items to evaluate adjacently, but if most items have already been fixed in the canvas, only few nodes will compose the input. This means that the worst function runs in amortized constant time, that is O(n) time for the worst case and O(c) in the average-case, which is the highest probable case. 
\\
So we can consider the total computational cost for the validate function as O(c) for the following reasons: if the worst function takes in input n items, then the treemap will be completed after a couple of instruction. The sum of the costs will be $2 \cdot O(c) +  \theta(1) \cdot 4 = O(c)$; if the worst function takes in input less than n items, it will take a greater number of iterations to complete the treemap, but each iteration will be less than O(n), and will slow down as the nodes will be fixed, costing still O(n).
\\
In this way, it is possible to evaluate the effort needed to extend the Squarified technique. The original control flow changes only in the else statement, where the validate function is added. Assuming that the Squarified algorithm has a computational cost of O(n), we can say that, the Squarified+ algorithm costs $O(n) + O(c)$, which is still O(n).

\section{Comparison tests}
Comparison tests have been developed to evaluate empirically Squarified+ effectiveness. Tests were designed to assess the heuristic's goodness in generating better treemaps, considering as main parameters the aspect ratio and the rectangles' homogeneity. We want also detect drawback effects if any, and what is the improvement volume. We already checked the computational cost needed in the previous section.

\subsection{Testing Process}
The testing process is made up by many test cases, one per each input size. Chosen sizes are 10, 50, 100, 500, 1000, 2000, 3000, and 4000, in order to simulate real cases study for "small" and "big" input data. The upper limit of 4000 nodes is due only to 8GB of memory available of the used machine. Moreover, we have used a virtual canvas of 1920 x 1080 pixels as screen where treemaps were arranged. 

Every test cases has been repeated 500 times, and each one follows these steps:
\begin{itemize}
	\item{First, it generates an input having a fixed size and leaves having random weight values.}
	\item{Then, it runs both Squarified and Squarified+ algorithms on the input;}
	\item{At last, it exports both trees' metrics to a file, in order to process them later.}
\end{itemize} 

The input generation process produces trees having a fixed size (e.g. 500 nodes) and depth 1. Leaves are given random weights values, ranging from 1 to 1 million. 
When the treemapping algorithms complete their executions, every node of both trees refers to a rectangle object, which will have appropriate height and width in according to the canvas size. \\
At this point, metrics can be calculated. Assuming \textit{AR} is a function that calculates the aspect ratio of a generic rectangle, by dividing its biggest side by the smallest one, reci is a generic rectangle object referred by a node i, and w is the weight of a node. We calculate:

\begin{itemize}
	\item{Mean Aspect Ratio:  $\overline{AR_{Tree}} = \frac{ \sum_{i=1}^{n}AR(rec_{j}) }{n}, node_{i} \in Tree   $;}\\
	
	\item{Weighted Mean Aspect Ratio:  $ weighted \overline{AR_{Tree}} = \frac{ \sum_{i=1}^{n}AR(rec_{j}) \bullet w_{i}  }{\sum{w_{i}}}, node_{i} \in Tree   $;}\\
	
	\item{Corrected Standard Deviation:  $ Std.Dev_{Tree} = \sqrt{\frac{\sum_{i=1}^{n} (AR(node_{i} - \mu))}{n-1}}   $;}
	
\end{itemize}

We introduced the weighted mean aspect ratio, beside the classic average calculation, in order to understand how much a treemap truly gets better, because in this way improvements to bigger nodes (with higher weights) become more relevant than improvements to smaller nodes. \\
Comparing two mean aspect ratio values we expect that the lowest value reflect the treemap having more squared items. We also expect that high variance will be symptom of non-uniform treemap's content, with squared and skinny rectangles, while low variance will indicate similar shaped rectangles, i.e. a more homogeneous treemap.

\subsection{Tests Results}
The table \ref{tab:tabResult} resumes the results of described tests. The success rate indicates how many times, in a test case, Squarified+ algorithms gives a better result for the considered metric, while mean improvement columns indicates the entity of obtained improvements.

\begin{table*}[]
	\centering
	\caption{Aggregated Tests Results \label{tab:tabResult}}
	\label{results}
	\begin{tabular}{|c|c|c|c|c|c|c|}
		\hline
		\textbf{\begin{tabular}[c]{@{}c@{}}500 Tests \\ per Size\end{tabular}} & \multicolumn{2}{c|}{\textbf{AR Tree}}                                                                                                   & \multicolumn{2}{c|}{\textbf{weighted AR Tree}}                                                                                          & \multicolumn{2}{c|}{\textbf{Standard Deviation}}                                                                                        \\ \hline
		\textbf{Tree size}                                                     & \textbf{\begin{tabular}[c]{@{}c@{}}Success \\ rate\end{tabular}} & \textbf{\begin{tabular}[c]{@{}c@{}}Mean \\ Improvement\end{tabular}} & \textbf{\begin{tabular}[c]{@{}c@{}}Success \\ rate\end{tabular}} & \textbf{\begin{tabular}[c]{@{}c@{}}Mean \\ Improvement\end{tabular}} & \textbf{\begin{tabular}[c]{@{}c@{}}Success\\  rate\end{tabular}} & \textbf{\begin{tabular}[c]{@{}c@{}}Mean\\  Improvement\end{tabular}} \\ \hline
		10                                                                     & 96,8\%                                                           & 5,53\%                                                               & 100\%                                                            & 4,02\%                                                               & 92,3\%                                                           & 5,1\%                                                                \\ \hline
		50                                                                     & 97,4\%                                                           & 2,22\%                                                               & 100\%                                                            & 1,55\%                                                               & 92,1\%                                                           & 5,0\%                                                                \\ \hline
		100                                                                    & 98,5\%                                                           & 1,89\%                                                               & 100\%                                                            & 1,13\%                                                               & 92,1\%                                                           & 6,1\%                                                                \\ \hline
		500                                                                    & 100\%                                                            & 1,48\%                                                               & 100\%                                                            & 0,62\%                                                               & 97,2\%                                                           & 16,5\%                                                               \\ \hline
		1000                                                                   & 100\%                                                            & 0,98\%                                                               & 100\%                                                            & 0,45\%                                                               & 96,0\%                                                           & 19,6\%                                                               \\ \hline
		2000                                                                   & 100\%                                                            & 0,64\%                                                               & 100\%                                                            & 0,33\%                                                               & 96,2\%                                                           & 27,0\%                                                               \\ \hline
		3000                                                                   & 100\%                                                            & 0,59\%                                                               & 100\%                                                            & 0,31\%                                                               & 94,7\%                                                           & 26,7\%                                                               \\ \hline
		4000                                                                   & 100\%                                                            & 0,44\%                                                               & 100\%                                                            & 0,24\%                                                               & 97,7\%                                                           & 54,0\%                                                               \\ \hline
	\end{tabular}
\end{table*}

\begin{figure}[h]
	\centering
	\label{meanAR}
	\includegraphics[width=1\linewidth]{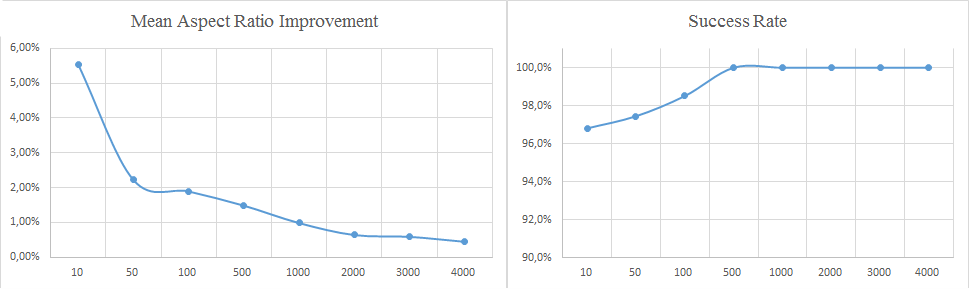}
	\caption{Squarified+ improvements of mean AR \label{fig:figure5}}
\end{figure}

\begin{figure}[h]
	\centering
	\label{wmeanAR}
	\includegraphics[width=1\linewidth]{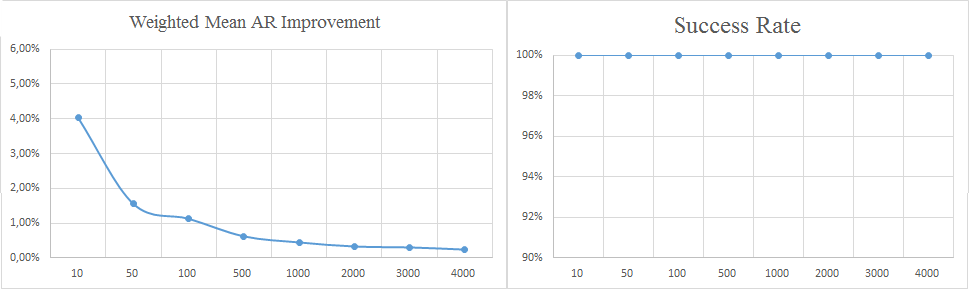}
	\caption{Squarified+ improvements of weighted mean AR\label{fig:figure6}}
\end{figure}

Analyzing the results, we can observe that aspect ratio improves mostly for small inputs; in fact, treemaps with 10 nodes have been improved more that 5\% in average. This result comes up in 97\% of tests, while the remaining 3\% ca. reveals an aspect ratio degraded in comparison to the Squarified technique.\\
However, as size increases (Figure \ref{fig:figure5}b) the percentage of success goes quickly up to 100\%, but the mean improvement  for aspect ratio slows down (Figure \ref{fig:figure5}a) due to higher treemap's density. In fact, dense treemap tends to uniform their rectangles, even if weights are distant, because the canvas has to be divided by a huge number of nodes, flatting weights differences.\\
Failures depend mostly by the worsening of latter nodes in the treemap. When the Squarified+ algorithm inverts the drawing direction, more space is given to the current row, so next nodes will have less. This usually does not have a strong impact, except for last items. In those cases, they could be arranged in a worse manner in comparison to how they would be placed by Squarified. For this reason, we use the weighted mean aspect ratios, so degradations to the smallest nodes does not affect the global treemap improvement.\\
This assumption does not change the Squarified+ results proportions, since the mean improvement remains almost the same for small and big inputs (Figure \ref{fig:figure6}a), but this time the success rate tell us that treemaps get better in any test (Figure \ref{fig:figure6}b).

\begin{figure}[h]
	\centering
	\label{stddev}
	\includegraphics[width=1\linewidth]{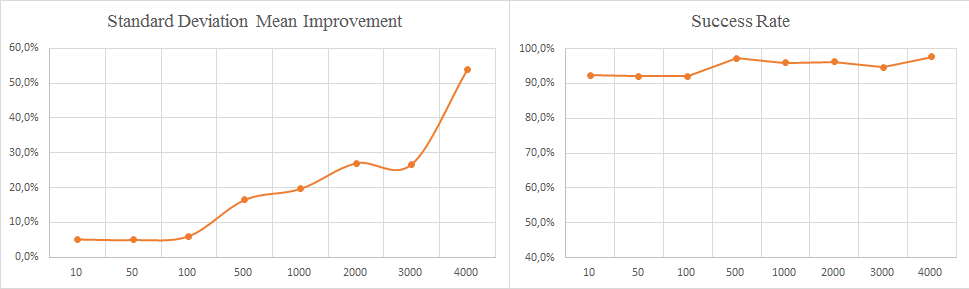}
	\caption{Improvements of standard deviation\label{fig:figure7}}
\end{figure}

Looking at Figure \ref{fig:figure7}, we notice that, similarly to the aspect ratio, standard deviation values enhance in most of the cases. Improvements also go up significantly together with input sizes. This is due to the less physic space given at size increasing. When items have small rectangles, even a few pixels of difference make the increment going up quickly. However, a non-improved standard deviation does not mean that Squarified+ has failed. Assume treemaps T1 and T2 as result of respectively Squarified and Squarified+ executions. Assume also that $AR(T1) > AR(T2)$ but$ SD(T1) < SD(T2)$, where \textit{AR} is the treemap mean aspect ratio and SD is the treemap standard deviation. In this situation, T2 standard deviation could be higher than T1 because T1 has only skinny rectangles, thus it would have a better homogeneity factor than T2, which has both elongated and squared rectangles. \\
\\
At last, we present some aggregated data, which confirm the good results of Squarified+. They include median values for aspect ratio and standard deviation metrics, and maximum values for the same metrics, to understand how much our heuristic can affect the treemap building process.
These data give a high-level overview about previous tests results. Looking at Table \ref{tab:tabMax}, we can see differences between median values of Squarified executions and Squarified+ ones (indicated by measurements denoted with "+"). As we expected, the median values for Squarified+ is always lower for each test case. On the other hand, in spite of low improvement results given by average calculations seen before, maximum values show good improvements to aspect ratios, which reach the 20\% for small inputs, and around 10\% for medium sized inputs. Standard deviation max improvements also give important results. For small inputs, treemaps' homogeneity get better about 50\%, while for big inputs such values go rapidly up to 150-200\%.

\begin{table*}[]
	\centering
	\caption{My caption \label{tab:tabMax}}
	\label{my-label}
	\begin{tabular}{|c|c|c|c|c|c|}
		\hline
		\multicolumn{4}{|c|}{\textbf{Median Value}}                       & \multicolumn{2}{c|}{\textbf{\begin{tabular}[c]{@{}c@{}}Max Improvements \\ Squarified+\end{tabular}}} \\ \hline
		\textbf{AR} & \textbf{AR+} & \textbf{Std Dev} & \textbf{Std Dev+} & \textbf{AR+}                                      & \textbf{SD+}                                      \\ \hline
		1,253       & 1,209        & 0,541            & 0,502             & 19,22\%                                           & 79,72\%                                           \\ \hline
		1,168       & 1,150        & 0,248            & 0,232             & 9,12\%                                            & 45,2\%                                            \\ \hline
		1,132       & 1,111        & 0,182            & 0,168             & 8,60\%                                            & 56,4\%                                            \\ \hline
		1,064       & 1,050        & 0,096            & 0,084             & 8,48\%                                            & 196,6\%                                           \\ \hline
		1,046       & 1,037        & 0,079            & 0,063             & 4,32\%                                            & 163,3\%                                           \\ \hline
		1,032       & 1,026        & 0,070            & 0,050             & 2,27\%                                            & 163,9\%                                           \\ \hline
		1,026       & 1,021        & 0,059            & 0,043             & 1,22\%                                            & 164,7\%                                           \\ \hline
		1,023       & 1,018        & 0,073            & 0,045             & 1,16\%                                            & 215,8\%                                           \\ \hline
	\end{tabular}
\end{table*}

\subsection{Validity Evaluation}
In this section, we discuss the threats to validity that could affect our results, focusing the attention on internal, construct, and conclusion validity threats.

\subsubsection{Construct Validity}
Construct validity threats can be due to random tree generation process performed during comparison tests. This was mitigated thanks to a high test repetition factor, which is 500 per each tree size and assigning random weights ranging from 1 to 1 million. These allow avoiding partial or non-exhaustive results, where input trees fit perfectly our extension and ensure a good distribution of weight values. 

\subsubsection{Internal Validity}
A strong internal validity threat that could have been present in this experiment refers to the treemaps' quality measurement in comparisons tests. Such threat was mitigated by defining a proper measurement that takes care about measurement problems in treemaps, which is the weighted mean aspect ratio. We are confident that other measurement would not have truly reflected the perceived changes in treemaps. 

\subsubsection{Conclusion Validity}
A definition of conclusion validity could be the improvement degree reached by our heuristic is real. With regard to our experiment, proper tests were performed to pragmatically demonstrate our hypothesis. Cases where Squarified+ results appeared equal or worse that Squarified results were explicitly mentioned and faced. 

\section{Conclusion}
In this work, we have developed a heuristic to extend the Squarified technique for graphing hierarchical data structures by treemaps, called Squarified+. Our algorithm is able to improve the process of arrangement of adjacent rectangles into the free space available, aiming to yield treemaps with better aspect ratio and homogeneity values, but without adding substantial computational complexity. To aim this, further optimization steps have been involved while items are being processed, allowing evaluating alternative ways of placing rows into the canvas and decide a better arrangement. In order to prove empirically that our extension achieves these goals, some comparison tests have been carried out, analyzing when and how Squarified+ gives enhanced treemaps compared to the Squarified method. Tests results confirm that the heuristic is able to give better treemaps, achieving more square-like and similar shaped rectangles. Despite the entity of the enhancement is revealed not very impressive, max values registered seem encourage the adoption of our new approach.\\
Currently, an aspect that emerges from our studies concerns about the absence of relationships between aspect ratio of parent nodes and their inner treemap's aspect ratio. Follow-up studies could examine if the used criteria could affect treemaps nesting process, aiming to spot any relationship among parent-children's aspect ratios and other key factors. \\
In conclusion, the effort put in this work is worthwhile due to the research of always better and more effective techniques for data representation. Effectiveness and efficiency are strict requirements nowadays, at most in human-computer interaction field, given that an increasing number of users is approaching to newest technologies, and to new graphical devices that offer always higher resolutions and quality \cite{treemap22}.

\bibliography{paper}
\bibliographystyle{plain}

\end{document}